\documentclass[12pt]{iopart}

\usepackage{iopams} 
\usepackage{graphicx}
\begin{document}

\title[Electric field filamentation]{Electric field filamentation and higher harmonic generation in a very high frequency capacitive discharges}

\author{Sarveshwar Sharma$^{\dag}$, N. Sirse$^{\ddag}$, A. Sen$^{\dag}$, J. S. Wu$^{\delta}$ and M. M. Turner$^{\ddag}$}

\address{$^{\dag}$ Institute for Plasma Research (IPR) and HBNI, Gandhinagar-382428, India}
\address{$^{\ddag}$ School of Physical Sciences and NCPST, Dublin City University, Dublin 9, Ireland}
\address{$^{\delta}$Department of Mechanical Engineering, National Chiao Tung University, Hsinchu, Taiwan}

\ead{nishudita1628@gmail.com}
\begin{abstract}
The effects of the discharge voltage on the formation and nature of electric field transients in a symmetric, collisionless, very high frequency, capacitively coupled plasma are studied using a 
self-consistent particle-in-cell (PIC) simulation code. 
At a driving frequency of 60 MHz and 5 mTorr of argon gas pressure, the discharge voltage is varied from $10$V to $150$ V for a fixed discharge gap. It is observed that an increase in the discharge voltage causes filamentation in the electric field transients and to create multiple higher harmonics in the bulk plasma. Correspondingly, higher harmonics, up to $7^{th}$ harmonic, in the discharge current are also observed. The power in the higher harmonics increases with a rise in the discharge voltage. The plasma density continues to increase with the discharge voltage but in a non-linear manner, whereas, the bulk electron temperature decreases. Meanwhile, the electron energy distribution function (EEDF) evolves from a Maxwellian at lower discharge voltages to a bi-Maxwellian at higher discharge voltages.
\end{abstract}
\submitto{\JPD}
\maketitle
\section{INTRODUCTION}
Capacitive discharges operating in a very high frequency (VHF) regime, 30-300 MHz, are now becoming popular for plasma processing applications including plasma etching and plasma enhanced chemical vapour depositions (PECVD). The benefits of using VHF are twofold; 1) as the driving frequency increases, the fractional electron power absorption increases and therefore at a constant power higher plasma densities can be obtained which increases the plasma processing rates, 2) at the higher driving frequencies, the sheath voltage drops and hence the sheath width decreases which reduces the ion bombardment energy toward the substrate. These benefits offer greater flexibility to process small-scale features without damaging the substrate. Furthermore, the shape of the electron energy distribution function (EEDF) also gets modified with a rise in the driving frequency and consequently affects  various electron-neutral collision processes and thereby the flux of reactive species on to the substrate.

Along with the benefits of higher processing rate and lower substrate damage, the electron power absorption in VHF driven plasma discharges also differs from those observed in traditional 13.56 MHz CCP discharges. In a collisionless regime, the electron power absorption from the rf field in a 13.56 MHz is mostly due to stochastic process \cite{IEEE_1988_16_638,IEEE_1998_26_955,SPTP_1972_16_1073,JAP_1985_57_53,PRL_2002_89_265006}. Stochastic heating in CCP discharges has been widely studied and extensively reported in the literature  \cite{Kawa_13_2006, Sharma_2013, Sharma_46_2013, Goz_87_2001, Turner_42_2009, Sharma_22_2013,
 Surendra_66_1469_1991, Sharma_24_2015}. Turner \cite{PRL_1995_75_1312} showed that pressure heating also plays a dominant role in sustaining collisionless discharges. Further studies by Schulze \textit{et al.} \cite{PRL_2011_107_275001} demonstrated that the drift ambipolar heating mode dominates the ionization in strongly electronegative discharges. A local electric field reversal during the expanding and collapsing phase of the sheath is also found to be an additional source of electron heating  \cite{JPAP_2008_41_105214,POP_2013_20_073507}. In contrast to 13.56 MHz driven CCP discharges, for VHF driven CCP devices the generation of an electron beams close to the sheath edge and which travels through the bulk plasma without collision and interacts with the opposite sheath is capable of causing ionization and sustenance of the discharge \cite{PSST_2015_24_024002,POP_2016_23_110701,PSST_2010_19_015014}. The electron beam is also responsible for the generation of strong electric fields in the bulk plasma \cite{PSST_2010_19_015014}. Such fields have been observed in both electropositive and electronegative \cite{PRL_2011_107_275001,POP_2016_23_110701} plasmas, and are found to exist in both symmetric and asymmetric CCP discharges  \cite{PRL_2011_107_275001,POP_2016_23_110701}. These electric field transients produce  bulk heating of the plasma and also drastically enhance the plasma density. There have also been observations of multiple electron beams which are found to be dependent on the driving frequency  \cite{POP_2016_23_110701}. It is further revealed that for a combination of discharge voltage and driving frequency, an independent control of ion flux and ion energy can be achieved \cite{POP_2018_25_080705}.

Another interesting feature of VHF CCP devices is the presence of higher harmonics of the applied frequency in the current and voltage characteristics of the discharges. Using a B-dot probe, Miller \textit{et al.} \cite{PSST_2006_15_889} detected the presence of higher harmonics in the electric field of the bulk plasma at a driving frequency of 60 and 176 MHz. Upadhyay \textit{et al.} \cite{JPDAD_2013_46_472001} performed numerical simulations of electromagnetic wave phenomena in an asymmetric CCP excited at 60 MHz and showed the presence of higher harmonic content in the electric field up to the 20th harmonic of the excitation frequency. The above studies revealed that the presence of higher harmonics could be one of the factors responsible for the non-uniformity in the plasma excited at VHF. More recently, Wilczek \textit{et al.} \cite{PSST_2018_27_125010} simulated voltage and current driven asymmetric CCP discharges in the low pressure regime. Their simulation results showed a curious asymmetry in the generation of harmonics, namely, that for a voltage driven device higher frequency components appeared in the rf current whereas for a current driven device no such harmonics appeared in the voltage characteristics.  The presence of higher harmonics in the rf current also revealed an enhanced excitation of plasma series resonance which further led to a drastic increase in the plasma density \cite{PSST_2018_27_125010}. Most of the studies mentioned above were performed in asymmetric CCP discharges. In view of the significant contribution of 
these transient electric fields and their higher harmonics in the dynamical properties of the VHF CCP devices it is important to carry out further explorations of their nature and dependencies on various device parameters in the VHF regime. In the present work, we carry out a detailed numerical simulation study of the effect of the discharge voltage on the electric field transients and the plasma parameters including the EEDF in a symmetric CCP discharge excited at VHF. Our simulations, carried out for a symmetric CCP device, show interesting features that link the appearance and characteristics of transient electric fields with the applied voltage and also delineate the corresponding changes in the EEDF that influence plasma kinetic processes. 

The paper is organized as follows. Section \ref{simulationScheme} provides a description of the simulation technique that is based on Particle-in-Cell/Monte Carlo collision (PIC/MCC) methods \cite{Birdsall_1991,Hockney_1988}. The physical interpretation and discussion of the simulation results are given in section \ref{simResults}. Finally, a brief conclusion and summary remarks are given in section \ref{conclusionDiscuss}.

\section{SIMULATION SCHEME AND PARAMETERS}
\label{simulationScheme}
\begin{figure}[htp]
\center
\includegraphics[width=8.5cm]{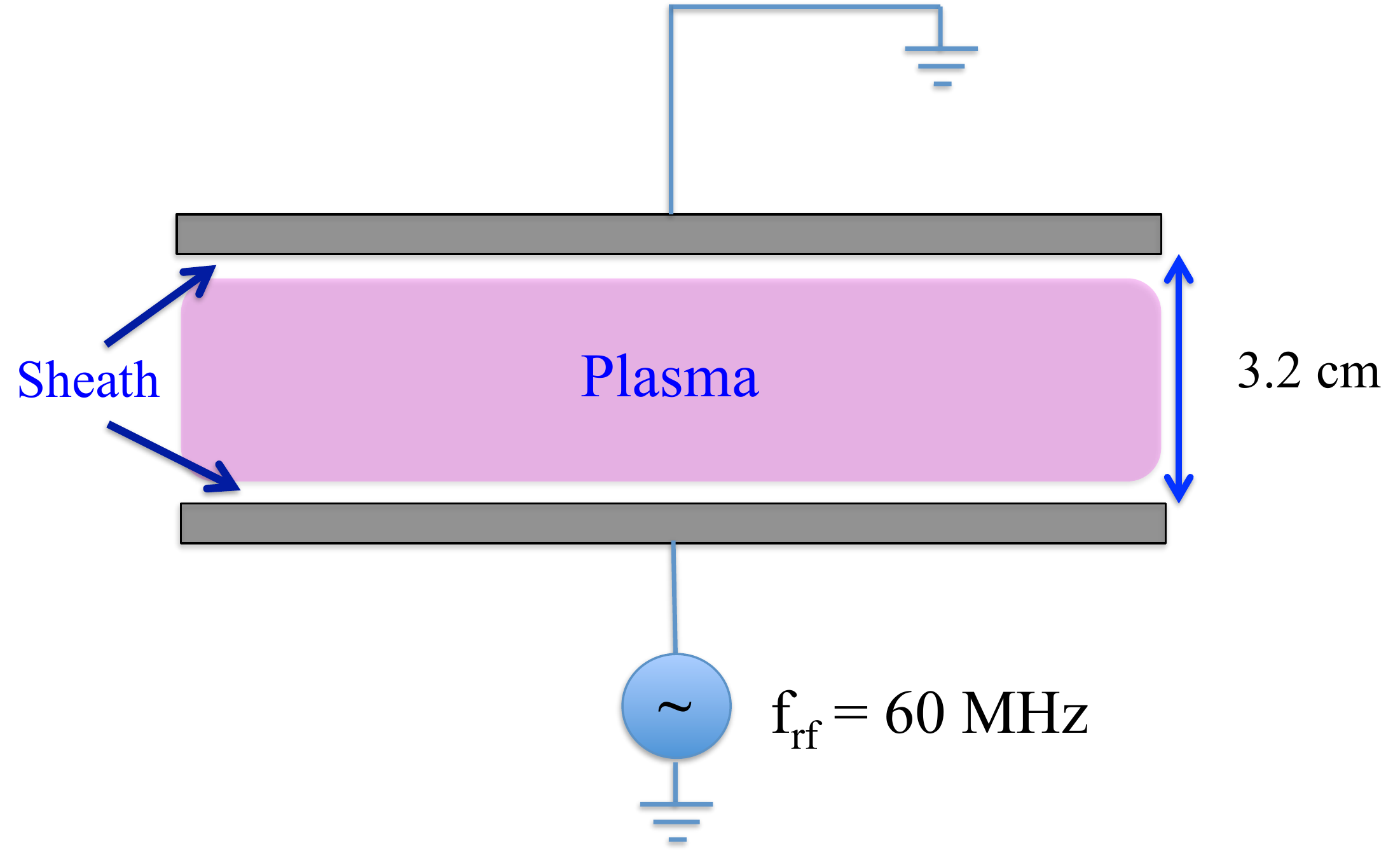}
\caption{Schematic diagram of voltage driven CCP discharge operated at $60$ MHz, $5$ mTorr pressure and for different voltages in an Ar plasma.}
\label{fig:figure1}
\end{figure}

A schematic diagram of a CCP discharge is shown in Fig.(\ref{fig:figure1}). We have simulated a voltage driven single frequency capacitive discharge in argon with a well-tested 1D3V, self-consistent, electrostatic, particle-in-cell (PIC) code. The simulation technique is based on Particle-in-Cell/Monte Carlo collision (PIC/MCC) methods \cite{Birdsall_1991,Hockney_1988}. The PIC simulation offers full information about the plasma kinetics such as the interaction of electrons with oscillating sheaths and includes the trajectories and loss of kinetic energy due to electron absorption at the walls etc. The code has been developed by Professor Miles M Turner at Dublin City University, Ireland and has been utilized in several earlier research works \cite{POP_2016_23_110701, PSST_2013_22_055001, PSST_2013_22_045004, PSST_2004_13_493, JPDAP_2004_37_2216, PRL_1996_76_2069, POP_2017_24_013509}. A comprehensive description about the simulation technique can be found in the literature \cite{PPCF_2005_47_A231, POP_2013_20_013507}. All important particle-particle interactions like electron-neutral (elastic, inelastic and ionization) and ion-neutral (elastic, inelastic and charge exchange) collisions are taken into account for all sets of simulations. However, processes like multi-step ionization, metastable pooling, partial de-excitation, super elastic collisions and further de-excitation are not considered here for simplification. In the simulation, the production of metastables (i.e. $Ar^*$, $Ar^{**}$) has been considered though we did not track them for output diagnostics. An appropriate choice of the spatial step size (i.e. smaller than the Debye length) and temporal step size (i.e. less than the electron plasma frequency) have been considered to take care of the accuracy and stability criterion of the numerics. We assume that the electrodes are planar and parallel to each other with infinite dimension and can be operated in both current and voltage driven modes. In our present analysis, we have chosen to operate in the voltage driven mode. Both the electrodes are perfectly absorbing for electrons and ions and the secondary electron emission is ignored for the present case. The gap between the electrodes is fixed at $3.2$ cm and the simulation region is divided into $512$ numbers of grids. We choose $100$ number of particles per cell for all sets of simulation. The background neutral gas is distributed uniformly having a temperature similar to that of ions i.e. $300$ K. One of the electrodes is grounded (x = 0) and an RF voltage with the following waveform drives the other one ($x = L$, where $L$ is the electrode gap):
\begin{equation}
\label{equ1}
 V_{rf}(t) = V_0 sin(2\pi f_{rf}t + \phi).
\end{equation}
To achieve steady state, the simulations were run for several hundreds of rf cycles. 

\section{SIMULATION RESULTS AND DISCUSSION}
\label{simResults}
\begin{figure}[htp]
\center
\includegraphics[width=16cm]{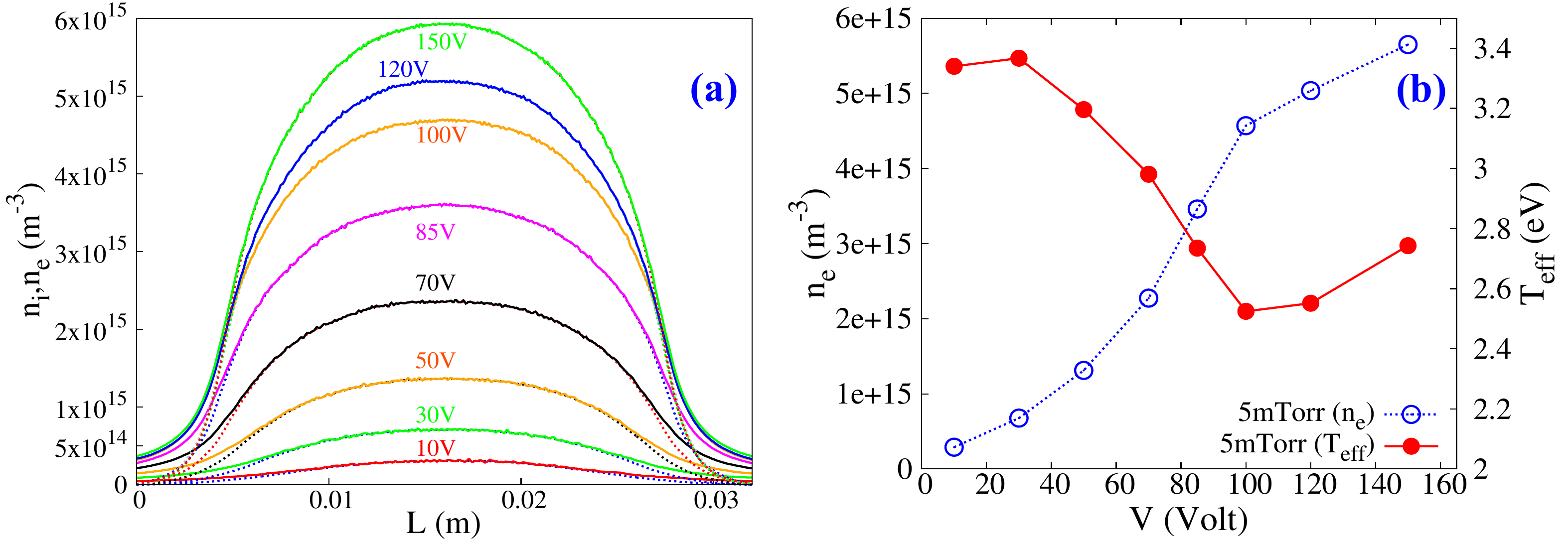}
\caption{(a) Plot of the time average electron (dotted curve) and ion densities (solid curve) at different discharge voltages. (b) Electron density ($n_e$) and effective electron temperature ($T_{eff}$) versus discharge voltage at the centre of the discharge for a constant driving frequency of $60$ MHz.}
\label{fig:figure2}
\end{figure}
In this section, we present our simulation results highlighting the role of the discharge voltage on the electric field transients and plasma parameters including the electron energy distribution function (EEDF). The applied rf frequency is $60$ MHz. At this driving frequency, we had previously observed strong electric field transients in the bulk plasma \cite{POP_2016_23_110701, POP_2018_25_080705}. Furthermore, this frequency is well above the transition frequency where the transients emerging from one sheath interact with the opposite sheath \cite{POP_2016_23_110701, POP_2018_25_080705}. The discharge voltage is now varied from $10$ V to $150$ V at a constant gas pressure of $5$ mTorr. The discharge gap is fixed at $3.2$ cm. 

Figure (\ref{fig:figure2}) (a), shows the time average electron and ion density profiles at different discharge voltages. The electron density at the centre of the discharge and the effective electron temperature is shown in figure (\ref{fig:figure2}) (b). The effective electron temperature is estimated from the electron energy distribution function (EEDF) using  $T_{eff}=\left(2/3  \right) \left( \int \varepsilon F(\varepsilon) d\varepsilon /\int  F(\varepsilon) d\varepsilon  \right) $, where $F(\varepsilon)$ is the self consistent EEDF and $\varepsilon$ is the electron energy. The EEDF is described later in the manuscript (figure 7). In a low pressure regime, which is the present case, the ohmic heating in the plasma can be neglected and therefore a classical inhomogenous plasma model predicts a linear increase in the plasma density with an increase in the discharge voltage i.e.
\begin{equation}
\label{equ2}
 n_{s} = \omega^2V_{rf}/\varepsilon_c
\end{equation}
where $n_s$ is the sheath edge plasma density, $\omega$ is the applied frequency, $V_{rf}$ is the discharge voltage and $\varepsilon_c$ is the collisional energy loss per electron-ion pair that is created. Our simulation results clearly demonstrate that this is not the case. Instead, a non-linear increase in the plasma density is observed. At the centre of the discharge, the electron density is $2\times10^{14}$ $m^{-3}$ at $10$ V which increases up to $\sim 6\times10^{15}$  $m^{-3}$ at $150$ V, i.e. an increase of $\sim 30$ times in the electron density is observed for $15$ times increase in the discharge voltage. A non-linear trend of plasma density versus discharge voltage was previously observed in both simulation \cite{JAP_1995_78_6441} and in experiments \cite{Wood_1991} predicting a different scaling law. The simulation results attributed this effect to more complex nature of the discharge such as three-dimensional effect which is not captured by one-dimensional PIC \cite{JAP_1995_78_6441}. According to equation (\ref{equ2}), a decrease in the value of $\varepsilon_c$ with discharge voltage might explain a higher increase in the plasma density, since $\varepsilon_c$ is a function of electron temperature, which is also changing in our case. As shown in figure (\ref{fig:figure2}) (b), the electron temperature is first decreasing from $3.4$ eV at $10$ V to $2.5$ eV at $100$ V and then increasing slightly up to $2.8$ eV at $150$ V discharge voltage. Thus the value of $\varepsilon_c$ will increase \cite{PSST_2001_10_76} which also does not support two times higher increase in the plasma density. These results clearly suggest the presence of a higher order phenomenon, which might play a significant role for enhancing the plasma density.  
\begin{figure}[htp]
\center
\includegraphics[width=16cm]{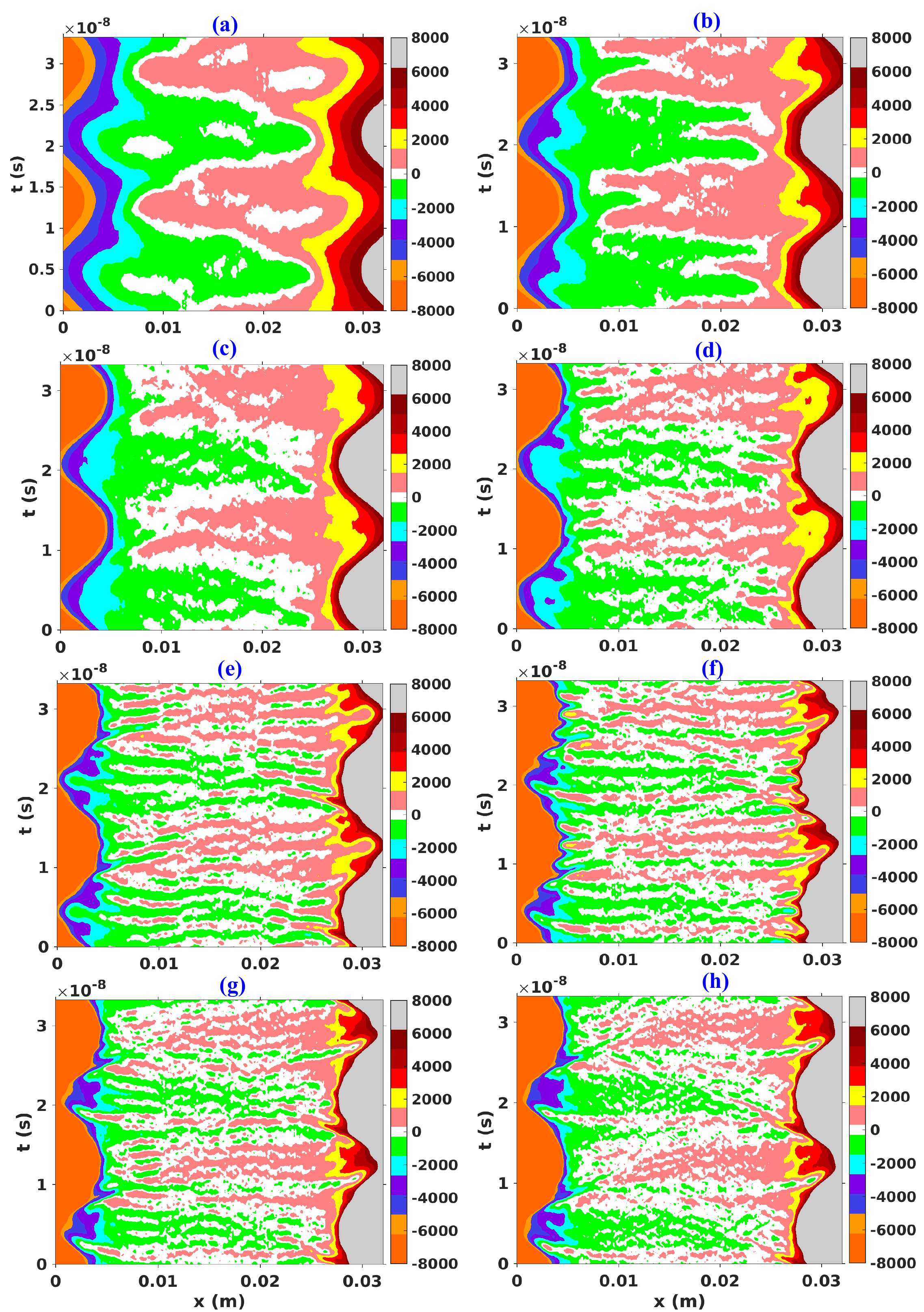}
\caption{Spatio-temporal evolution of electric field in the discharge at (a) $10$ V, (b) $30$ V, (c) $50$ V, (d) $70$ V, (e) $85$ V, (f) $100$ V, (g) $120$ V and (h) $150$ V discharge voltage.}
\label{fig:figure3}
\end{figure}
We next turn to the electron heating in order to understand the non-linear plasma density behavior. Figure (\ref{fig:figure3}) shows the spatio-temporal profile of the electric field for $8$ different discharge voltages (from $10$ V to $150$ V) and for the last $2$ RF cycles. As shown in figure (\ref{fig:figure3}) (a), at a discharge voltage of $10$ V, we observe strong electric field transients in the bulk plasma. These electric field transients have broad temporal widths and are spatially extended up to the opposite sheath. As the discharge voltage increases up to $70$ V, the transients start splitting in time and become thinner as seen in figures (\ref{fig:figure3}) (b) $-$ (d). At discharge voltages of  $85$ V (figure (\ref{fig:figure3}) (e)) and $100$ V (figure (\ref{fig:figure3}) (f)), the transients are very thin. After $100$ V, the transients start to merge again. In all cases, the transients are spatially extended such that they interact with the opposite sheath and modify the instantaneous sheath edge. This modification is strongest at $100$ V discharge voltage. It is also observed that after $100$ V discharge voltage, the nature of the sheath changes drastically when compared to the lower voltage cases i.e. the sheath is not collapsed fully. However, for other cases the sheath almost reaches its minimum at the electrode.
\begin{figure}[htp]
\center
\includegraphics[width=16cm]{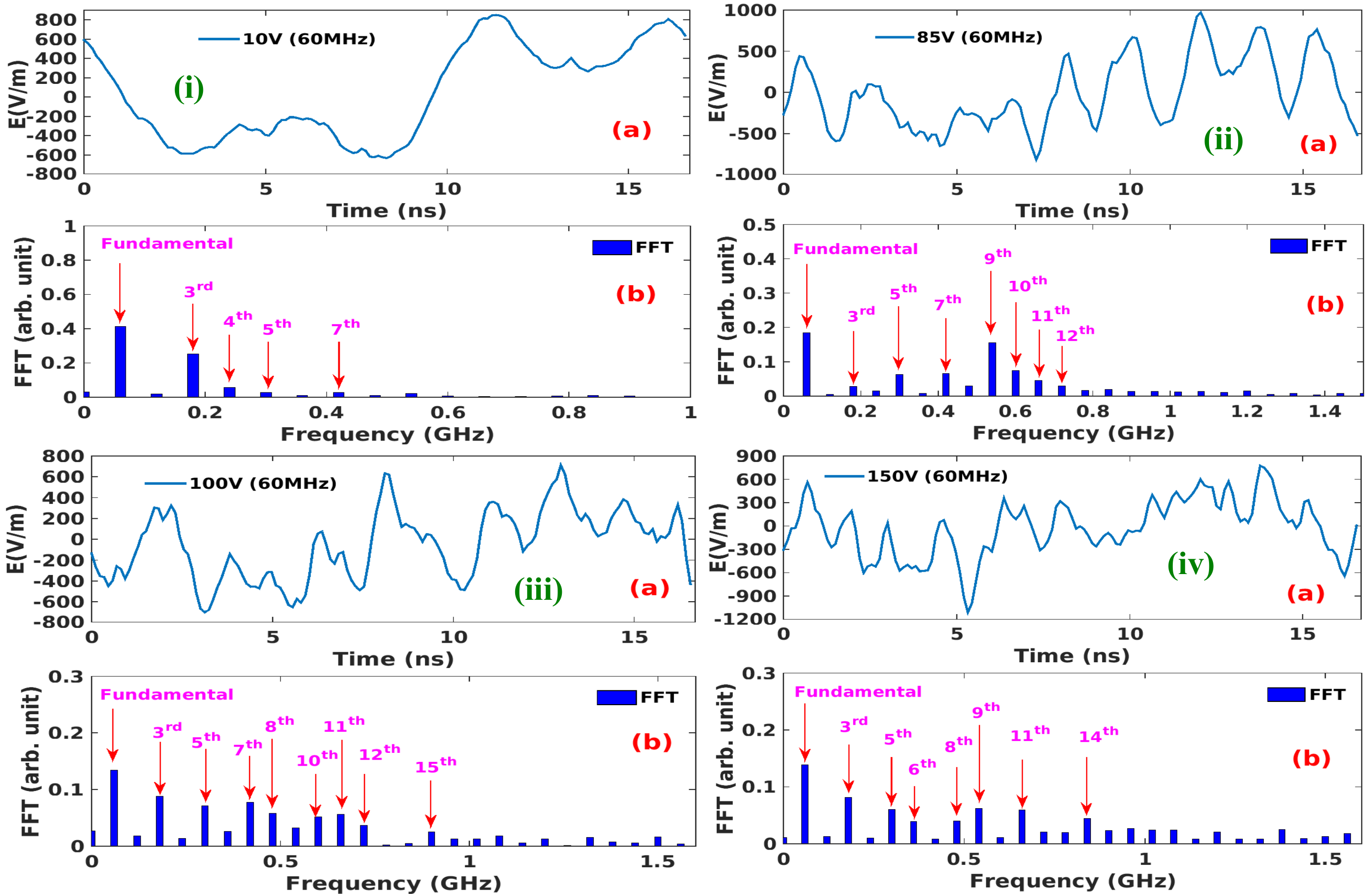}
\caption{Time variation of electric field at the center of discharge and its FFT at (i) $10$ V (ii) $85$ V (iii) $100$ V and (iv) $150$ V applied voltage for constant driving frequency, $60$ MHz.}
\label{fig:figure4}
\end{figure}

Figures (\ref{fig:figure4}) (i), (ii), (iii) and (iv) show the electric field versus the rf phase at the center of the discharge and its Fast Fourier Transform (FFT) in time for $4$ different discharge voltages \textit{i.e.} $10$ V, $85$ V, $100$ V and $150$ V respectively. For all the cases, we find a significant presence of electric fields at the centre of the discharge and also observe higher harmonics. The FFT shows that at $10$ V, fig (\ref{fig:figure4}) (i), two dominant frequencies are present \textit{i.e.} $\sim 60$ MHz and its $3^{rd}$ harmonic $180$ MHz. The electron plasma frequency, $f_{pe}=\sqrt{n_e e^2/\varepsilon_0 m_e}/2\pi$  at $10$ V is $0.1269$ GHz, which is more than $2$ times higher than $1^{st}$ harmonic and less than $3^{rd}$ harmonic. While increasing applied voltage up to $100$ V, we have seen that both, number of higher harmonics, and its amplitude keep increasing (see Fig (\ref{fig:figure4}) (ii) and (iii)). At $85$ V (Fig.(\ref{fig:figure4}) (ii)), the $540$ MHz is the most dominant frequency ($\sim 9$ times of applied RF frequency) after fundamental and is slightly higher than $f_{pe} = 536.3$ MHz. The $3^{rd}$, $5^{th}$, $7^{th}$, $10^{th}$, $11^{th}$, $12^{th}$ and $13^{th}$ harmonics are also present, however, their amplitudes are low in comparison to the fundamental driving frequency and its $9^{th}$ harmonic. In $60$ MHz case, using coupled plasma-EM wave simulation, Upadhayay \textit{et al.} \cite{JPDAD_2013_46_472001}  also predicted significant presence of the $9^{th}$ harmonic. At $100$ V (Fig. (\ref{fig:figure4}) (iii)), the dominant frequencies are $180$ MHz, $300$ MHz, $420$ MHz, $480$ MHz, $600$ MHz, $660$ MHz and $720$ MHz i.e. $\sim$ $3^{rd}$, $5^{th}$, $7^{th}$, $8^{th}$, $10^{th}$, $11^{th}$ and $12^{th}$ harmonic of applied RF frequency respectively, and their amplitudes are also significant. The plasma frequency in this case is $f_{pe} = 0.6137$ GHz which is lower than one of the dominant high frequency mode \textit{i.e.} $660$ MHz. At $150$ V, (Fig. (\ref{fig:figure4}) (iv)) the high frequency content has less representation compared to the fundamental driving frequency. Here $f_{pe} = 0.6173$ GHz and notice that the $11^{th}$ and $14^{th}$ harmonics are higher than $f_{pe}$; however, its contribution is less when compared to the fundamental driving frequency. It appears here that by increasing and decreasing the discharge voltage from $100$ V the transients are merging, and the contribution and numbers of higher harmonics are decreasing. 
\begin{figure}[htp]
\center
\includegraphics[width=16cm]{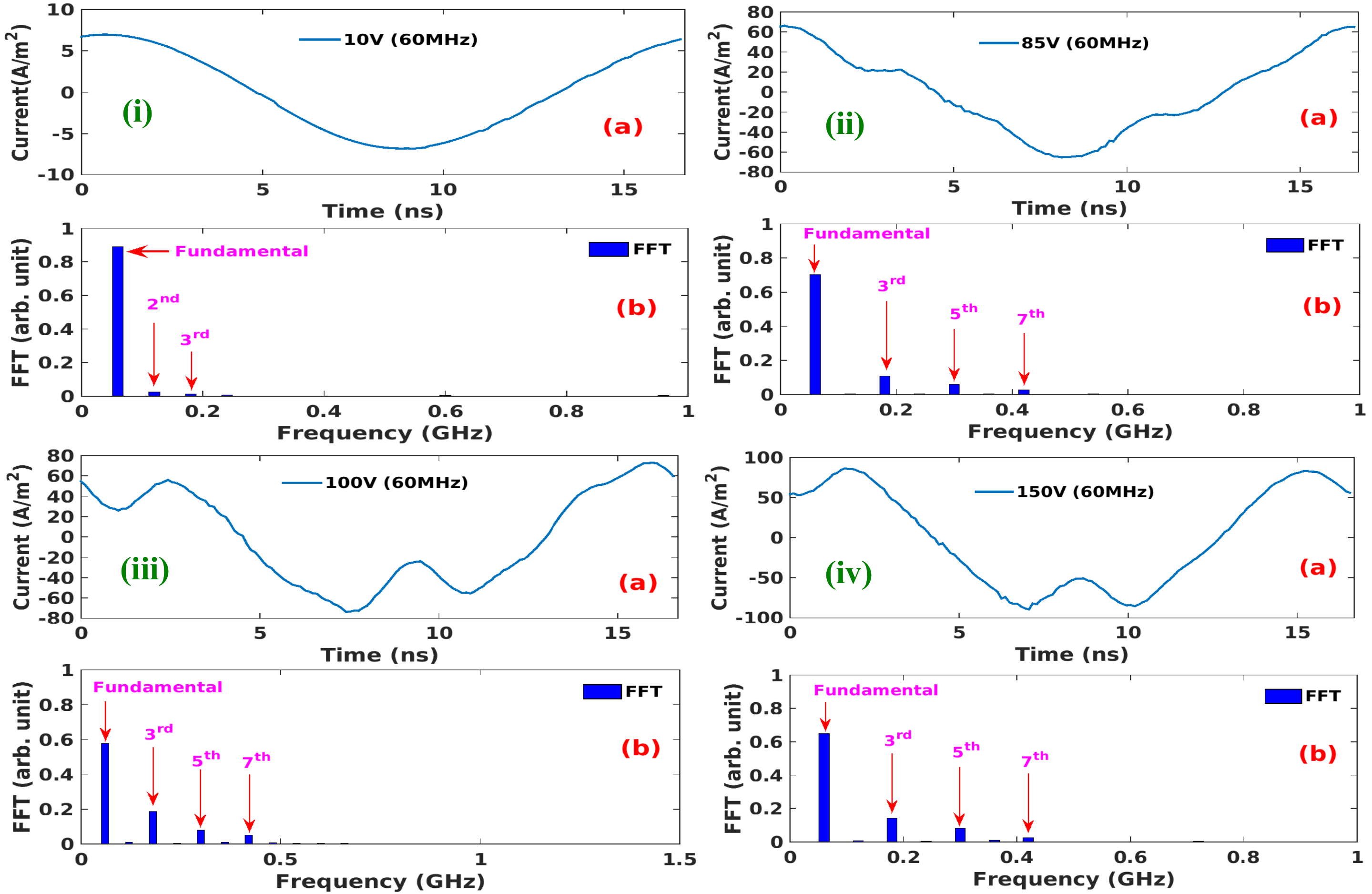}
\caption{Time variation of electric current density at the electrode and its FFT at (i) $10$ V (ii) $85$ V (iii) $100$ V and (iv) $150$ V applied voltage for constant driving frequency, $60$ MHz.}
\label{fig:figure5}
\end{figure}

The presence of higher harmonics in the discharge and an increase in their amplitudes is suggestive of a non-linear increase in the plasma density (as shown in figure (\ref{fig:figure2})) since they are more efficient in power deposition. The appearance of electric field transients and its higher harmonics in the bulk plasma are related to the energy gained by the electrons from the oscillating sheath which trigger bursts of high-energy electrons. The energy gained by the electrons from the moving sheath is dependent on the sheath edge position which in our case is decreasing from $\sim 10$ mm at $10$ V to $\sim 4.5$ mm at $100$ V and further increasing up to $5.5$ mm at $150$ V discharge voltage. An initial decrease in the sheath edge position up to $100$ V discharge voltage suggests a reduction in the sheath velocity which means that the electrons have significant time to interact with the moving sheath and therefore energy gained by the electrons increases. After gaining energy from one sheath, the fast electrons are then able to reach to the opposite sheath resulting in the modification of the instantaneous sheath position. Figure (\ref{fig:figure4}) shows that the modification of the instantaneous sheath edge position is highest at $100$ V discharge voltage which confirms that the beam velocity is higher in this case. The triggering of multiple beams of electrons is attributed to the back flow of bulk electrons due to space charge created by the first beam generation as described by Wilczek et al \cite{Wilczek_POP_23_2016_063516}. Furthermore, for per sheath expansion, the number of electron beam further depend on the electron response time and sheath velocity. In our case, the sheath velocity is decreasing up to 100 V discharges voltage and therefore the no of beam  increases. The flow of bulk electrons towards the expanding sheath edge causes non-sinusoidal current waveform at the electrode. Figure (\ref{fig:figure5}) shows the current density from simulation and their Fast Fourier Transform at the powered electrode for the same $4$ discharge voltages \textit{i.e.} $10$ V, $85$ V, $100$ V and $150$ V respectively. We observe that for all voltage cases the simulated current is non-sinusoidal and higher harmonics are present. The higher harmonics of current is negligible at $10$ V, however, their contribution increases with discharge voltages. The strength and contribution of higher harmonics in the current is highest at $100$ V discharge voltage  (fundamental contribution least i.e. $58\%$) which also corresponds to a strong modification in the instantaneous sheath edge position (see fig (\ref{fig:figure3}) (f)). Similar type of higher harmonics in current was observed recently in a simulation done by Wilczek \textit{et al.} \cite{PSST_2018_27_125010} for an asymmetric CCP driven in voltage mode case.
\begin{figure}[htp]
\center
\includegraphics[width=8cm]{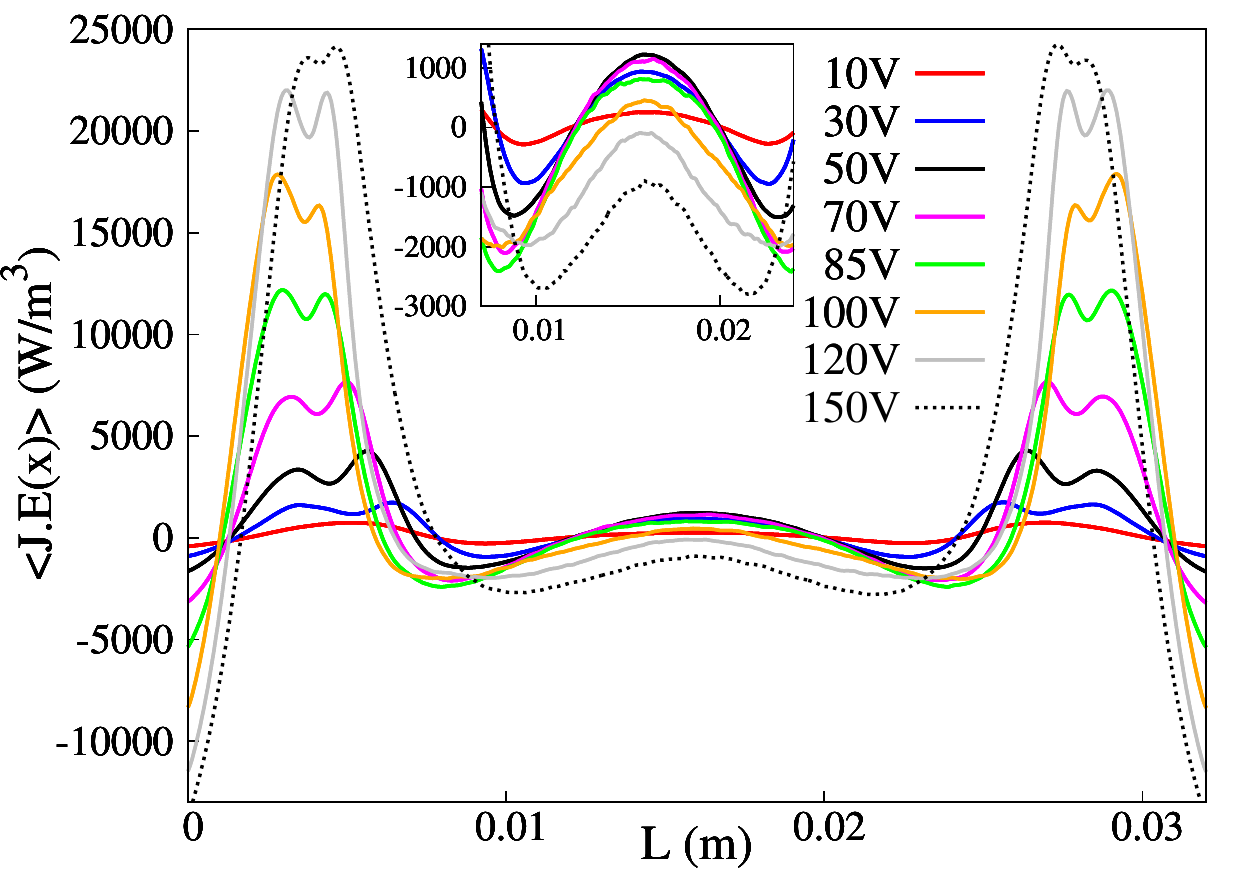}
\caption{Spatial profile of time-averaged electrton plasma heating ($\langle J.E \rangle$) for different discharge voltages from $10$ V to $150$ V.}
\label{fig:figure6}
\end{figure}

Figure (\ref{fig:figure6}) displays the spatial profile of time-averaged heating $<J.E>$ at different applied voltages from $10$ V to $150$ V. For all discharge voltages, the heating is maximum in the sheath vicinity and its magnitude increases with an increase in the discharge voltage. However, it is important to note that heating at the center of bulk is positive at $10$ V and its magnitude increases by rising voltage up to $70$ V ($1200$ $W/m^3$). After $70$ V, the heating slightly decreases, but is still positive at $85$ V, and then decreases further at higher voltages and becomes negative $<J.E>$ after $100$ V \cite{IEEETPS_2006_34_696, IEEETPS_1991_19_144}. This effect of electron cooling inside the bulk results from an electric field in the bulk operating in opposition to the fast electrons created from sheath, as well as phase-mixing between bulk electrons and tail electrons \cite{IEEETPS_2006_34_696, POP_2014_21_073511, CPP_2015_55_2015}. Physically this phenomenon may be explained as follows: the disturbances at the plasma-sheath boundary inject pulses of hot electrons into the bulk plasma. In the absence of electric fields, these pulses propagate into the bulk and produce some electron current density there that may be more or less than the total current density at the electrode. Since the current must be conserved, the electric field must adjust itself to produce a similar current in the bulk. This could lead to an electric field of either sign, and hence positive or negative bulk heating. As one changes the applied voltage, the phase of these pulses of hot electrons relative to total current in the bulk presumably changes. Hence, one might expect to find changes in the sign of the bulk heating as a function of applied voltage. Adjacent to the positive $<J.E>$ in the sheath edge is a region of strong electron cooling which is similar to the pre-sheath regime predicted by Kaganovich \cite{IEEETPS_2006_34_696}. These regions arise from the reflected, bunched electrons interacting with the bulk plasma. We believe this provides the energy to form the plasma polarization electric field and the oscillations within the bulk plasma.
\begin{figure}[htp]
\center
\includegraphics[width=8cm]{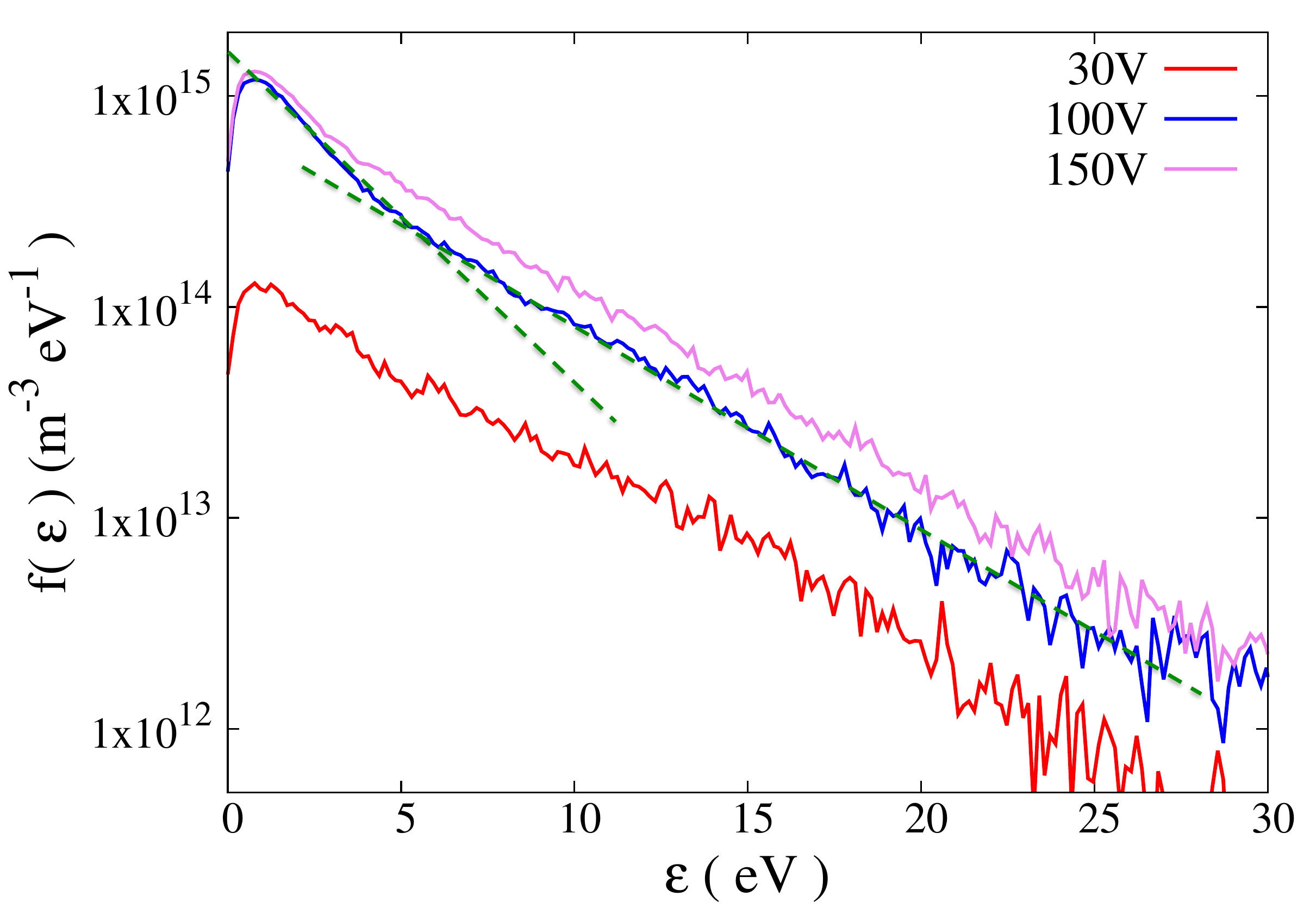}
\caption{ EEDF at the center of discharge at $30$ V, $100$ V and $150$ V discharge voltage.}
\label{fig:figure7}
\end{figure}

Finally, the effect of discharge voltage and higher harmonics generated in the bulk plasma on the EEDF is examined. Figure (\ref{fig:figure7}) shows the EEDF at the center of the discharge for $30$ V, $100$ V and $150$ V. It is observed that the EEDF is nearly Maxwellian at $30$ V and then turns into a bi-Maxwellian one at $100$ V, and finally becomes nearly Maxwellian again at $150$ V. This transformation of EEDF is attributed to the change in the nature of the electric field transients versus discharge voltage. At $30$ V, the electric field transients in the bulk plasma are thick in nature and the amplitude of higher harmonic is low in comparison to fundamental driving frequency. Furthermore, the instantaneous sheath edge position is not significantly modified.  This suggests that the beam of electrons are not effectively confined therefore do not create significant ionization and keep the density of low energy electrons low. Within the bulk plasma, the fast electrons are still redistributing their energy with the low and mid energy range electrons in the bulk plasma by some non-linear mechanism therefore producing a Maxwellian type distribution of electrons. As the discharge voltage increases up to $100$ V, the population of both low energy and high energy electrons keep increasing but the rate of increase of low energy electrons is greater when compared to high energy electrons. This changes the shape of the EEDF to a bi-Maxwellian type. This may be due to the better confinement of fast electron and significant contribution of higher harmonics which produces the low energy electron population due to ionization process. The modification of instantaneous sheath edge position further confirms that the fast electrons generated from one sheath are interacting with the opposite sheath during expanding phase and therefore further boosting the energy of fast electrons which is favourable for the ionization processes. After $100$ V, the population of high energy electrons are still increasing, whereas, increase in the population of low energy electrons are minimal. This turns the shape of EEDF at $150$ V to a nearly Maxwellian one. These results suggest that the rate of increase in the population of low energy electrons is reduced. This transition may be attributed to the merging of the electric field transients and therefore decrease in the number of higher harmonics and its amplitude in comparison to the fundamental driving frequency. Note that the instantaneous sheath edge position at $150$ V is not significantly modified which means that the confinement of fast electrons is not effective. From the above EEDF results, it can be concluded that the dominant presence of the higher harmonics (such as in the case of $100$ V discharge voltage) predict less energy dissipation of the energy of electron beam, which after confinement between 2 opposite sheaths produces ionization. This increases the population of low energy electrons and therefore drive a bi-Maxwellian type EEDF. At $10$ V and $150$ V discharge voltage, the lower presence of the higher harmonics predict higher energy dissipation of the electron beam energy and therefore does not produce effective ionization after confinement. The energy dissipation of the electron beam is a non-linear process and therefore an analytical model describing such an energy transfer is difficult and out of the scope of this work.

\section{SUMMARY AND CONCLUSION}
\label{conclusionDiscuss}
In summary, using PIC/MCC technique we have simulated a voltage driven symmetric CCP system of an argon plasma  driven at a VHF. In particular, the effect of the discharge voltage on the electron dynamics and plasma parameters including EEDF have been studied. A spatio-temporal analysis of the electric field  shows presence of electric field transients in the bulk plasma. These electric field transients are due to the presence of the multiple high energy electron beams which are generated from near the time-modulated sheath edge position. The electric field transients become temporally sharper as the discharge voltage is increased. The FFT of the electric field in the bulk plasma shows presence of higher harmonic contents of the frequency up to $20^{th}$ harmonic of the driving frequency. These higher harmonics are responsible for an efficient power deposition in the bulk plasma through non-linear interaction and also drive a non-linear increase in the plasma density with a rise in the discharge voltage. The higher harmonics contents up to the $7^{th}$ harmonic of the driving frequency are also observed in the current at the electrode. The time-averaged electron heating $<J.E>$ is positive in the bulk plasma up to $100$ V discharge voltage and thereafter negative electron heating \textit{i.e.} electron cooling is noticed. This is also reflected in the electric field transients which start merging again after $100$ V discharge voltage. The EEDF is found to be Maxwellian at low discharge voltage, whereas, it is bi-Maxwellian at $100$ V and then again become nearly Maxwellian type EEDF at higher discharge voltages. An increase and decrease in the higher harmonic contents is responsible for the transformation of the EEDF with the discharge voltage. The simulation results concludes that the generation of higher harmonics due to multiple beams of electrons is an effective mechanism for the power deposition in capacitive discharges. This leads to enhance the plasma density which assure an enhancement in the plasma processing rates. The transformation of EEDF to bi-maxwellian, \textit{i.e.} cold bulk and hot tail, concludes that the substrate damage could be lower due to reduced ion bombardment energy without affecting the plasma chemistry which is mostly governed by the hot electrons. 


\end{document}